# Computer of Things a Proposal to Speed up IoT Development

Alireza Ahadpour Shal[1], Amin Hjizadeh[2]

1- Faculty of Electrical Engineering and Information Technology, RWTH-Aachen University, Germany

2- Department of Energy Technology, Aalborg University, Esbjerg, Denmark

***Abstract:*** ***This paper propose and predict the need for a new line of computer production that can facilitate and accelerate the improvements of things and systems towards IoT networks. The proposed computer that is named Computer of Thing, CoT, will speed up the establishment of smart world in a synchronized way with similar standards, communication protocols, and hardware and software. Also, the need for a new standardization body in addition to those who exist is stated that will motivate governments to invest more on IoT development. However, to show the importance of proposed CoT, a brief review on IoT present status is performed and then some terminologies and classification of IoT networks is presented. This classification helps manufacturers to gain a proper view for their future products development. Then, observing from production point of view the need for CoT is highlighted. Finally, as smart systems may work in different dimensions of applications via internet or intranet connectivity, the basic specifications of proposed CoT are presented. Consequently, due to nature of CoTs, a new multi- dimension multi-hierarchy control strategy is proposed.***

## I. Introduction

In the next wave of industrial development in the era of industry 4.0 and internet of things, process, control and computing will be outside the realm of the traditional method. In the Internet of Things (IoT) paradigm, many of the objects that surround us will interact smartly using embedded information and communication systems in the environment around us.

Doing a literature review on IoT, demonstrates that most applied researches and visible improvements are reported after 2012. In most studies IoT is considered as comprised of a number of functional blocks to facilitate various utilities to the system such as, sensing, identification, actuation, communication and management. Not considering huge companies like IBM and Google, there exist many distributed academic literatures on the concept of internet, networking and smart system that are reported under IoT concept. IoT security [1,2], IP addresses [3], cloud services and computing [4,5], sensors networks [6], hardware and software for smart things [7-11], etc., are some of these unorganized researches. Some researchers had reviewed practical prototype implementations. For example, Stojkoska et. al. have reviewed a state-of-the-art smart home integration into IoT [12]. Smart cities evolution toward the internet of cities considering common obstacles has been addressed by Schleicher et.al, in which key needs have been identified and architectural guidelines for implementation have been presented [13]. Moreover, smart energy networks like a microgrid has been reported by [14, 15]. But the academic share in IoT development is very low in comparison with leading companies.

Leading companies like IBM, Google, Intel, Microsoft and Cisco are listed among top twenty companies, illustrated in Figure.1, in the IoT field [16]. For example Cisco introduced many new products mainly on network connectivity, IoX and Fog applications, data analytics, security - cyber and physical, management and automation, application platform. Mazak iSMART Factory, Cisco IoT Field Network Director, and Cisco Jasper platform are three real IoT based implementation reported by Cisco [17]. IBM has real solutions for factories, environment, sports, vehicles, buildings, home, and retail. Bluemix as cloud, Watson as IoT platform, Maximo as asset management, Tiririga as facility management, etc., are some examples of IBM products [18]. Google android of things (formerly known as Brillo) is a new comprehensive IoT platform for building smart devices on top of Android APIs and Google's own services. Google IoT cloud platform as well as google weave as a platform for IoT things communications are two other approaches reported by Google [19].

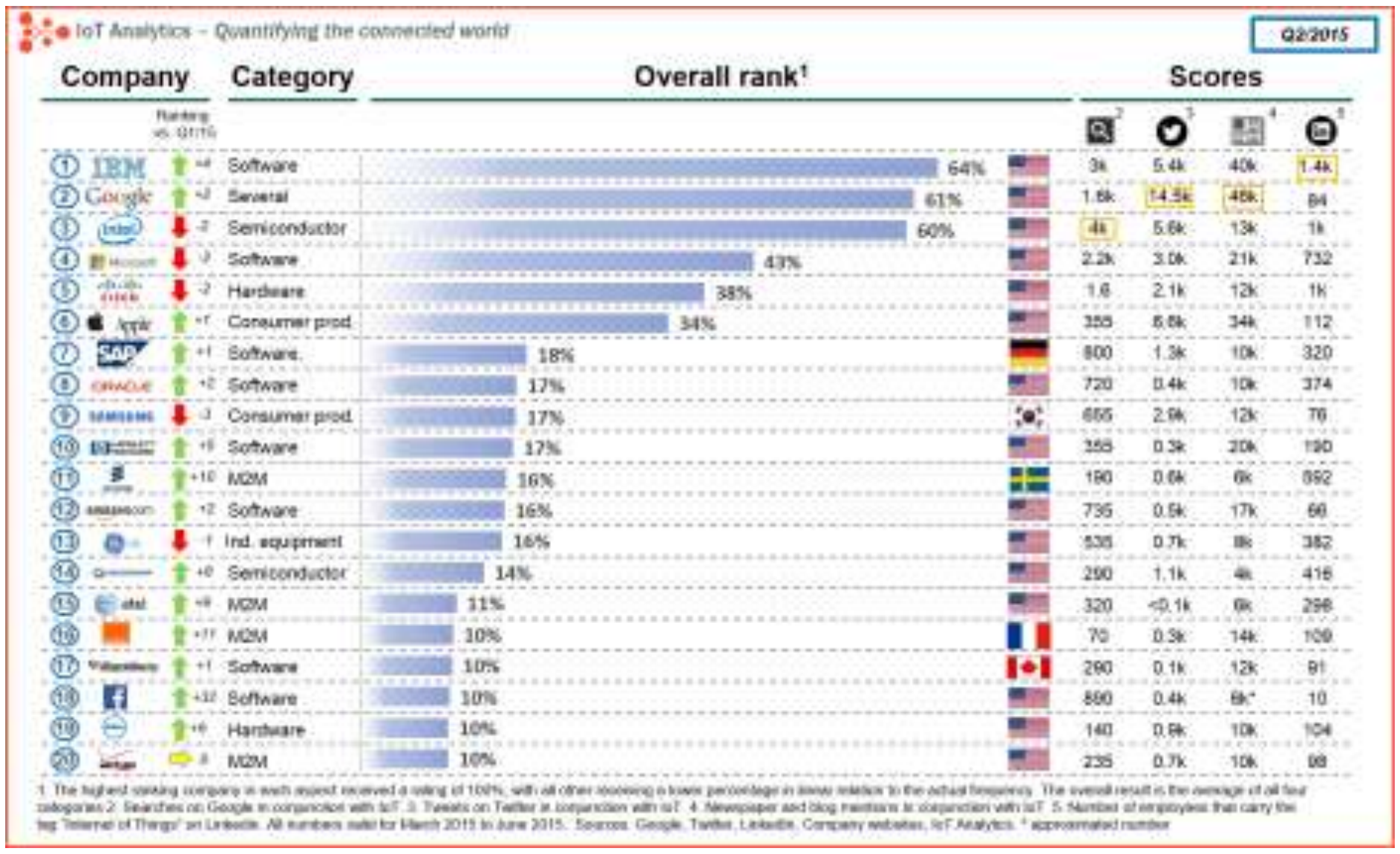

**Figure 1: Top 20 IoT companies [15]**

Figure.2 shows the main architecture that leading companies like Google and IBM Utilize to have IoT established. In this approach, 3 main stages can be obviously observed. 1- Ordinary things. 2- Evolution of things by computers or digital processing boards that can be connected to the internet or clouds. 3- Cloud or internet based processing.

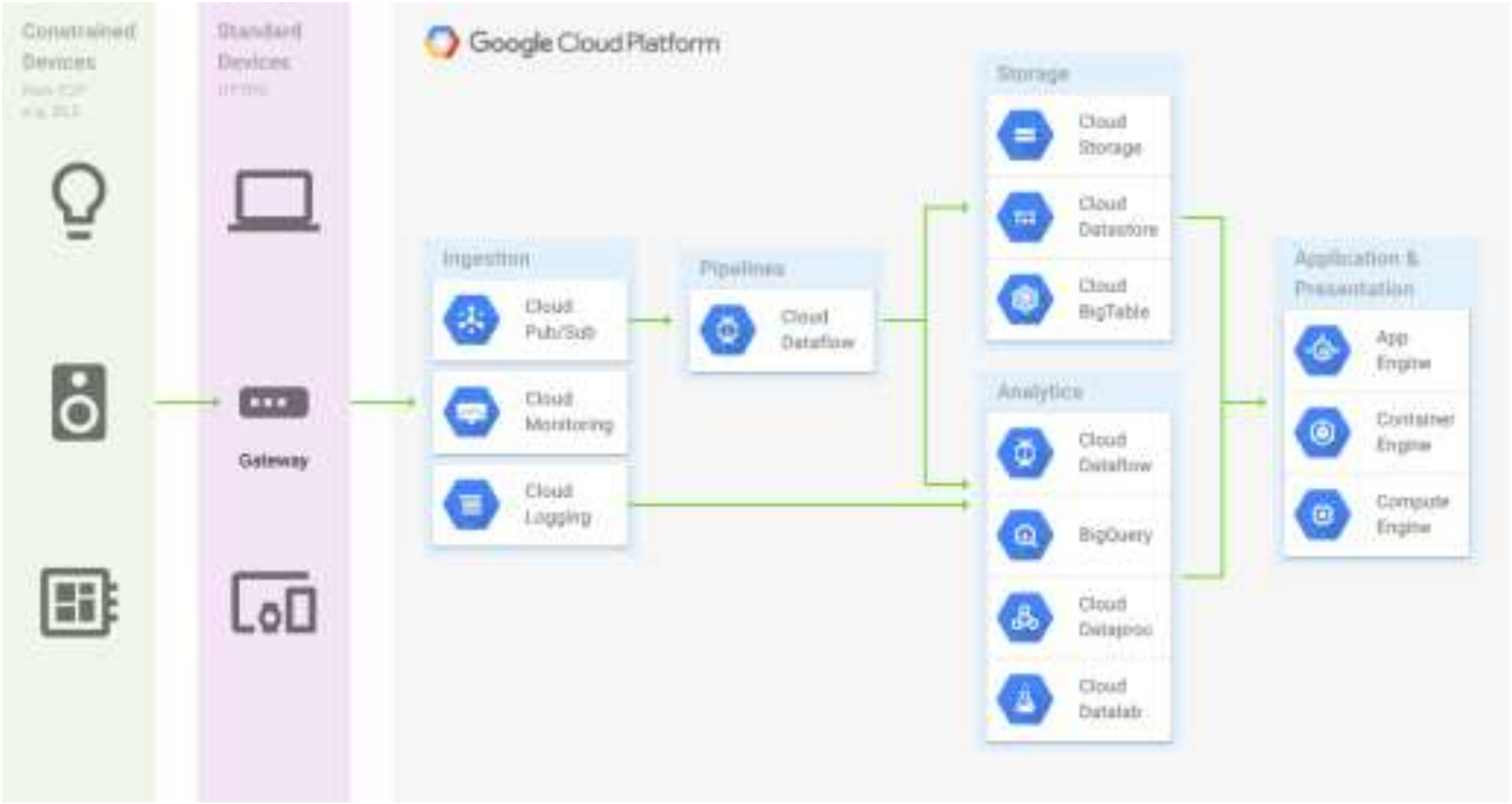

**Figure 2: IoT architecture based on google cloud platform [18].**

Based on reports the IoT evolution procedure is going on mostly by the map defined by pioneer companies. And most important data of IoT network will be process in cloud environment of limited companies and organizations. On the other hand, IoT is going to change organizational structure of all companies and organizations and their data exchange [20]. This fact may make governments reluctant to invest in this filed because of information privacy and security unless they have their own controlled clouds. Accordingly, IoT progress needs a comprehensive map defined by an internationally recognized official organization to regulate, organize, standardize and synchronize all researches and movement. This official organization would be responsible to set policies for all autonomy aspects of regional IoT networks to have their own security and privacy of information. This organization establishment would speed up IoT evolution and saves time, energy, money. Recently, there have been some organization that are trying to standardized some technologies related to IoT concepts like ALLJOYN, IIC, OIC, IoTGSI of ITU-T, IEEE P2413, [21], [22]. But all mentioned organizations do not have enough authority to manage the map of IoT evolution specifically in terms of regional autonomy and security of control and manage things and data.

It is also very important to give a suitable outlook of future compatible products to all manufacturer around the world. And it seems very essential to propose a classified approach that helps ordinary manufacturers to improve their products by their own policies using market available product that improve their system for IoT structure. Meanwhile, leading companies and their partners provide some processing boards, application platform, and solution for manufacturers to be connected to their exclusive IoT cloud. This approach should be seen in convergence of stage one and two in figure.2. But this approaches are not motivating enough for majority of manufacturer to produce their product for IoT compatibility.

Therefore, this work will try to simplify existing technology for improve things and systems toward IoT. Then it will suggest a new approach that can help manufacturer to be motivated to develop their product faster, easier and in a unique way towards IoT. Moreover, based on IoT trend, a proposal for a new line of computer, called computer of things, CoT, will be given in this paper. COT is highly recommended to be produced by leading computer or mobile industry to make construction of global IoT easier, faster and with the participation of manufacturers around the world

In the next section of this paper, based on surveys done on IoT, some terminologies with detailed definitions will be presented that provides a proper outlook for manufacturer to improve their products for IoT operation. Then two types of IoT network will be presented in section III, in which, the place and importance of special computer will be shown both in system and network level. Then in sections IV, based on smart systems operation in IoT networks, a new controlling strategy will be proposed. Finally, in section V, suggested CoT with some basic specifications will be presented.

# II. Terminologies and definitions

IoT is an emerging technology and many manufacturer do not have enough information about its trend. So it seems finding unified and simple terminology in this field will be essential. This paper will use following terms for this aim.

## *A. Computer of Thing (CoT)*

At the beginning of this paper, this term is the used for all kind of smart board or computers that are used to smartly control systems or intranet networks and make them compatible for IoT infrastructures by using some standard communication protocol and interfaces. But as it will be describe in section V, the aim of CoT in this paper is a proposal of new line of Computer production in different classes to have a classified approach to improve systems in an accelerated, unified, and organized way for IoT applications.

### B. *Thing (T)*

Thing (T) is simply anything that is aimed to be electronically equipped in order to be used in smart system and in turn in IoT network. It may be a simple artificial products such as a door or be a piece of thing in a complex system such as a heater in a car. It also can be a natural thing such as a tree, a plant, etc. Note that sometimes a sub-system in a complex system can be called thing.

### C. *Electronically Promoted Thing (EPT):*

Electronically promoted thing (EPT) is a T that is equipped with proper equipment in order to make it possible to be electronically connected to other EPTs or a processor in a smart thing (ST) or in a smart system. An EPT does not have embedded processor and should be controlled as a component within a ST or SN by a processor. Accordingly an EPT stays in lowest control hierarchy level in a system or network. The equipment installed on an EPT depends on the thing's importance and range of involved activities and usages. The equipment can be; sensors, actuators, different wire connectors (e.g. Some I/O ports in physical level that does not need software or processor for data exchange). An EPT can be fixed location or nomadic.

### D. *Smart Thing (ST)*

Smart thing (ST) is an EPT that is equipped with a local processor that controls and optimizes the ST's operation. Depending on complexity of the thing and the range of involved applications, a ST is equipped with suitable processor and memory (a simple CoT). Consequently, a ST can have the capability of having more standard input and output ports including ports that need protocol exchange, wireless connectivity, and internet connectivity.

Accordingly, a ST not only can control itself smartly but also can be connected to other EPTs and STs with in a smart system (SS). A ST connected to higher level of control hierarchy not only exchanges proper processed data but also gets the policy instruction for its control process. Note that a ST has just one processor and can be fixed location, nomadic and mobile. To have a simple illustration of T, EPT, and ST symbolic schematics of them are shown in figure.3.

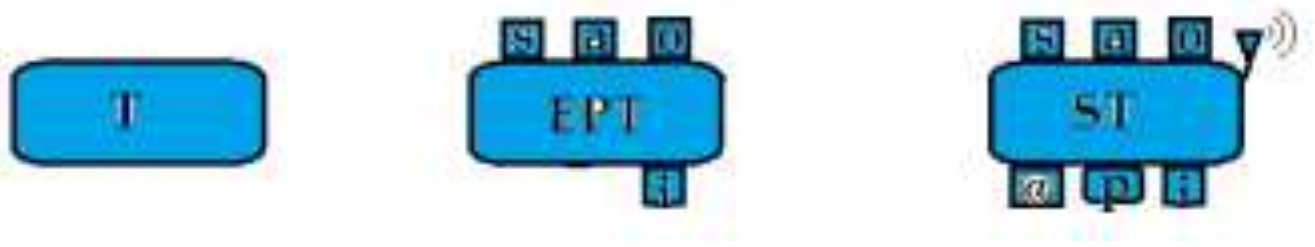

**Figure.3: T, EPT, and ST. (i & o: wire input & output connections or ports, P: equipped with suitable processor and memory, @: internet connectivity, : wireless connections)**

### E. *Smart System (SS)*

Smart system consists of some components including Ts, EPTs or STs. The components of a SS can be Main, important, or trivial depending on their functionality. Without main component SS will not operate, without Important components SS works but not safely and properly, and without Trivial components, the system can operate well but lacking some extra options.

A SS operates automatically and autonomously with specific objectives and can be fixed location, nomadic, or mobile. In IoT, an active SS in connected or controlled by at least one specific network of application in a time via connecting to higher level of SSs, CoTs, or internet based applications. A SS mostly have a ST (with a simple CoT) as its main controller with the highest hierarchy that manages all internal and external communications of SS. On the other words, main controller of SS not only manages and monitors other components of the system but also plays an important role to properly work in different networks.

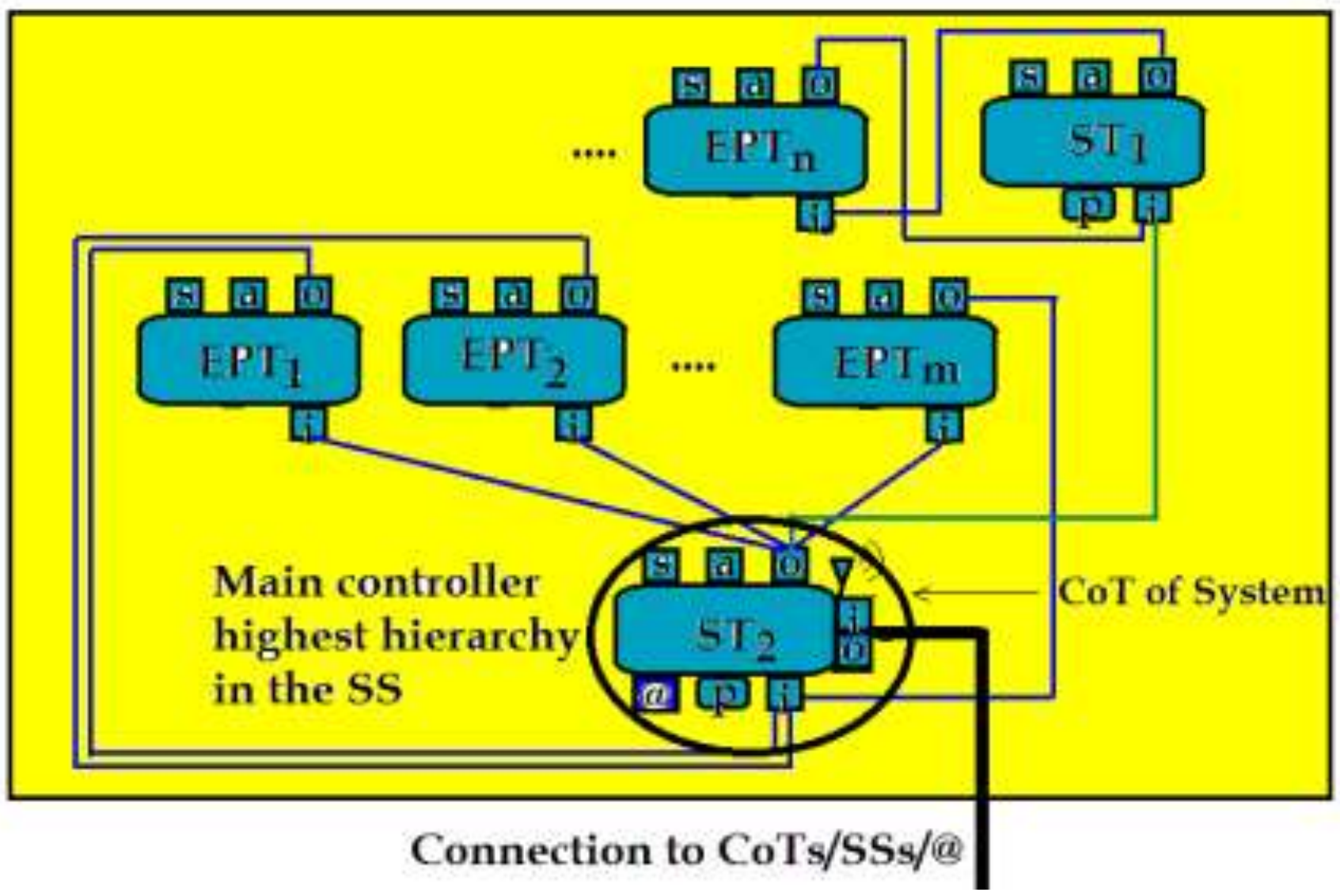

**Figure.4: an example of a smart system with n EPTs and two STs. One ST act as a main controller of the SS for connecting to other smart systems, internet of applications, or other CoTs in Physical network of things.**

Sometimes smart thing (ST) and smart system (SS) can be used interchangeably. But in fact SS is a more complex and mostly consists of at least one STs and other components. For instance, a smart engine within a car is a smart thing but the car that consist of smart engine and smart air conditioning etc., is called a smart system. On the other words, a SS have STs and also sub-SSs as its components. However, the distinguishing feature between SS and ST is that a SS can directly be a member of a network of application with an IP address. While, a ST can directly connected to a network of application it can be also consider as SS.

### *F. Smart Network (SN)*

A basic Smart Network (SN) is an interconnection of different smart systems, as members, controlled by a power full processor that can be called computer of thing. But network concept is more complex and can have many types and variety in terms of topology, size, scope, autonomy, locality, accessibility, security, application, etc. [23]. A smart network is developed for data interchange and analysis as well as controlling and monitoring aims to provide better services for the network or sub network members. SN definition is very similar to those of computer network with all network members, nodes, and services, sub networks etc. The same concept is applied in IoT based networks in which all members are smart systems or smart Things.

It is useful to mention that a smart network can have a fix or unfixed number of STs or SSs as its members and should have a dynamic ability of registering or deregistering members. In contrast with a ST that will not operate correctly if it loses some of its components, a network will properly work in case of its members' deregistering. In case of simple network a CoT provides all services for the members of the network. But network has more complex definition in which the service provider of the network is another network, an application based software via internet, a cloud computing center etc. However, the property and operation of a network depend strongly on its physical and logical topology and scope of operation.

# III. Types of Smart Network

Network concept is very complicated and different types of networks has been defined concerning different point of view. It would be very convenient for manufacturer to see networking ability of their SS products based on product point of view. They should know how SS should be produced to establish desired IoT connections and how to operate smartly in all data exchanges.

Accordingly, it is needed to categories network based on SS connection to IoT infrastructure concerning to location of operation. With this point of view, two types of networking can be imagined in IoT networks. First type is centralized network or location based network that is mostly called intranet, in which a SS works via local connection of a specific place. In such case a company or organization provides services only for members in the vicinity or location. Second type is location free and mostly internet based network that can be called decentralized network. Providing services in this type of network is not dependent on the location of operation and each member can use the services everywhere connected to the internet or global network of IoT.

An example of centralized network is a car parking lot network in a city center that wants to provide a smart network for its customers and provide them with a proper security, payment method, and desired parking space based on parking time, etc. This local network has its own way of communication with smart cars in the location and will not provide this level of service for smart cars out of location. The operation and connection management is done by CoT of the SN in cooperation with its members that are smart cars ad SSs. Some other location based networks or smart things can be find in location of smart homes and smart microgrid. Figure.5 shows a location base network with a CoT as main processor of the system that not only controls the members but also makes connection with other networks. It should be noted that the communication method among members and between members and CoT can be designed in different way depending on the nature, level of security, types of service, specifications, and features of the SN. It should be mention that a parking lot network itself can be a member of a decentralized network that provides some level of information for car out of location via internet of things facilities.

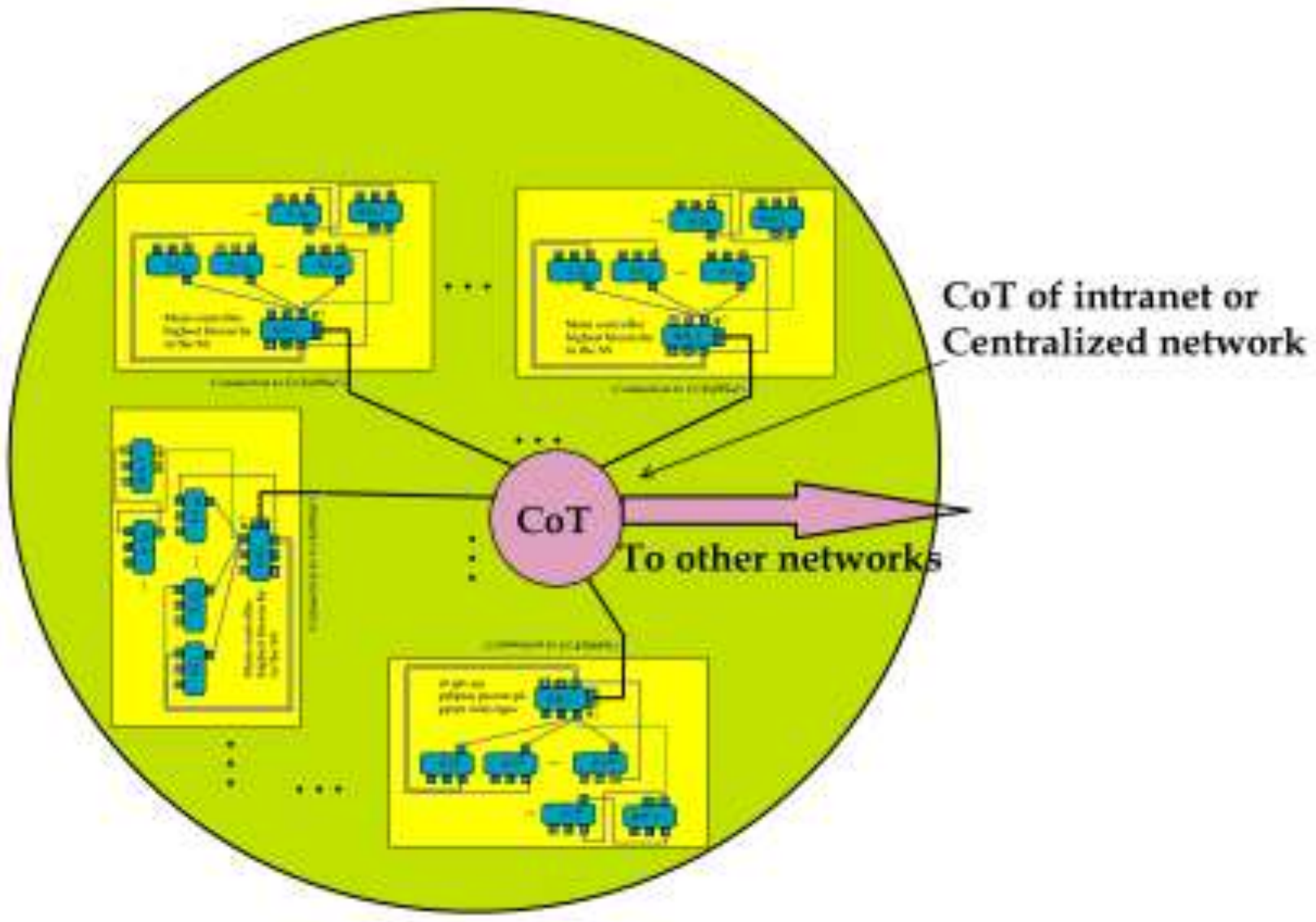

**Figure.5: A centralized SN which is location based and is equipped with a central CoT. The hierarchy in the network can be seen from the connections provided from ST up to SS and then to higher control levels.**

Decentralized network is a network in which members have no limit of distance or location to be a registered member. Smart system can exchange information via internet structures using specific application software to use a specific service or information. For example, suppose that there is an internet based application called “emergency services” work on some SSs installed on human body or on a car. The installed SS have a proper hardware and operating system (CoT) that can install this application in advance. When an emergency situation happens the smart system will be connected automatically to the internet and send critical information such as heart attack, accident, patient situation to the service provider via application software. The server of the application will smartly inform the related and closest doctor, ambulance, and hospital (all are connected to the same application directly or indirectly) about the accident. Then healthcare and emergency services will be provided in the shortest time and optimized way for the injured person. Some other examples of decentralized network can be find in [24], [25].

However, a centralized network can be a member of a decentralized network in which some internet base applications update their information based on local network information. For example in a centralized network such as a “hospital local network” update all its local information and then send a portion of them (like ambulance, and doctors information) to a decentralized network of application (e.g. the mentioned application that was called “emergency service”).

Is should be noted that the scope of this work is not IoT infrastructure and related topics like backbone and servers of IoT, IoT protocols, security, cloud and data analysis, etc. this work tries to show how system can be improved to be compatible to work in future IoT infrastructure.

Figure.6 show three decentralized networks in which the members are from different places. Each network is formed based on a specific application of operation. As it can be seen, network of application one (Ap1) has 3 SSs as members from different locations.  With the same principle, Ap2 has two STs and one SS as its members, and Ap3 has two various local SNs as its members. The hierarchy of each member and types of services provided in location free internet based application will be defined in software of application that is not shown in the picture.

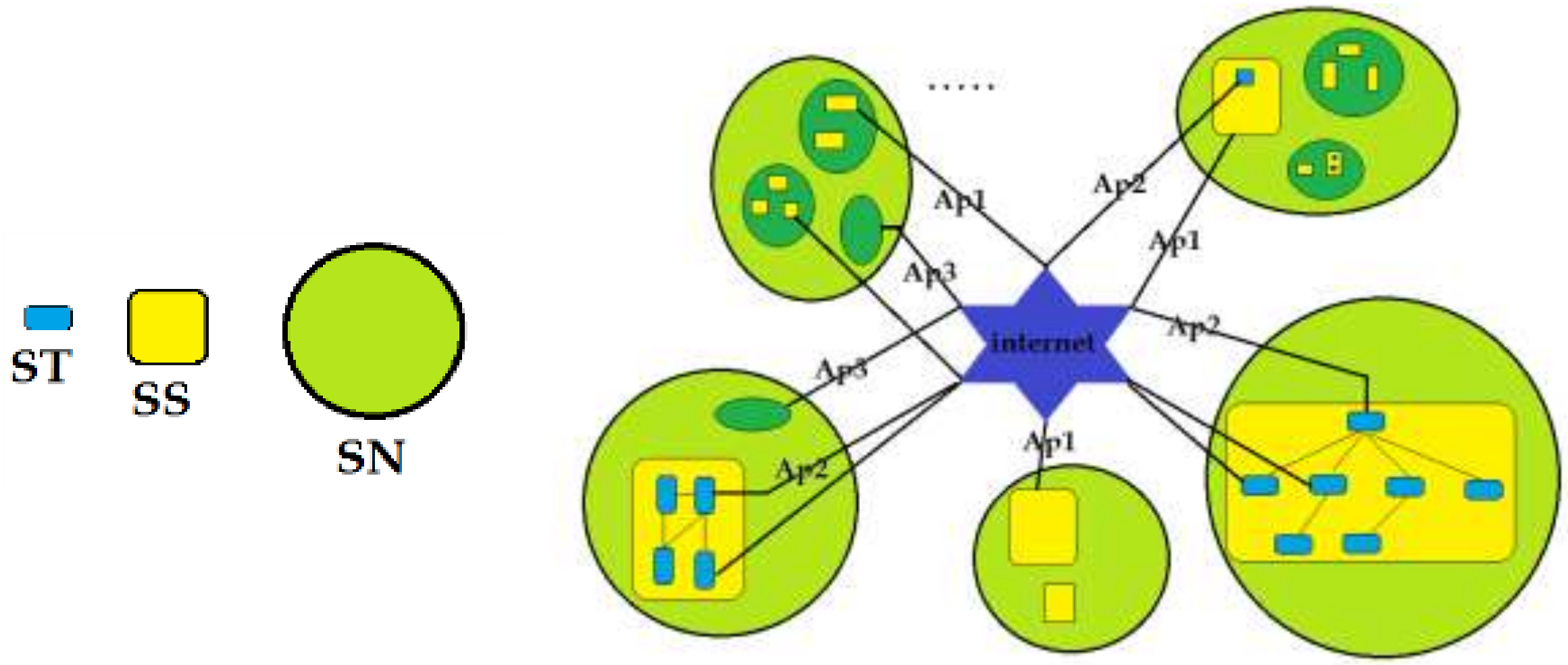

(a) (b)

**Figure.6: (a): Abbreviated graphic symbols for ST, SS, SN. (b): decentralized network that is location free and internet based network. A decentralized network registers its members via a specific software of application. Three network is shown for Ap1, Ap2, Ap3.**

Each network of application (e.g. AP1) is design for specific types of services or operation called "dimension of application". As mentioned, a SS may works in different dimension of applications. Depending on hierarchy of each network, the smart system will have a specific hierarchy place in each network. In order to properly program the processor or CoT of ST, SS or SN, the hierarchy in each dimension of application should be defined (Figure.7).

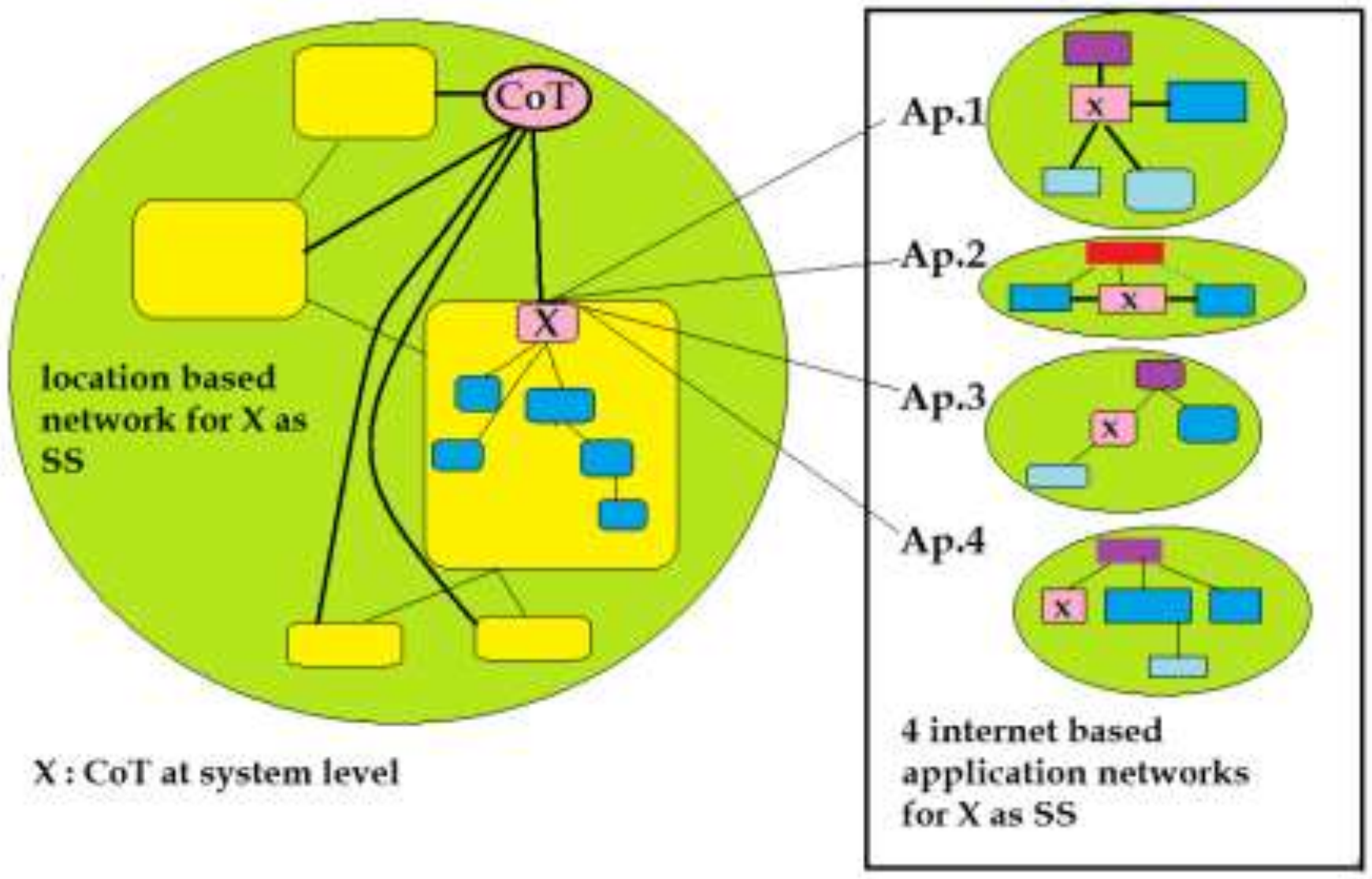

**Figure.7: A smart system as a member of different decentralized networks**

Just in case of proper hierarchy definition in each application the main processor of a system would be able to operate properly in all connected networks. This leads to suggestion of Multi-Dimensional Multi-Hierarchy Control method will be described in following section.

Considering complex nature of networking, some SNs may be a member of another higher level SN and act as a SS member of a higher level SS. Based on what is mentioned it is very important for a CoT of a SS or SN to be able to work in both centralized and decentralized networks. Accordingly, a CoT require at least some capabilities that will be described in the section V.

## IV. Proposed Multi-Dimensional Multi-Hierarchy Control

During recent years there have been a vigorous urge to have automatic controlled and intelligent systems. Most of this intelligent systems as a final industrial product relay on an embedded system with a programmable chip. Smart systems tend to have automatic and multi optional operating methods that are plug and play in the sense that they connect to and operate with other desired STs, SSs, and SNs in different level of control and communication.

In reality, a specific network consists of different local and sub-local networks with a specific pattern of connections and hierarchies (multi-level of hierarchy). A specific system may work in one or more dimension of applications in which each application requires a specific type of operation and data exchange in specific network. So that a SS may belong to more than one network of application, simultaneously. In each dimension of application network depending on the system level, it has its own hierarchy. The same pattern exists in IoT based smart systems and networks described in section III.

Also a system can be autonomous, partially-autonomous and non-autonomous. An autonomous system is a system with some targeted objectives and priorities that has no constrains and controls from other systems of higher hierarchy. The optimize operation of an autonomous system is defined just based on its objectives. A partially-autonomous system is a system with some targeted objectives and priorities that has some constrains and controls from other systems of higher hierarchy. The optimize operation of a partially-autonomous system is defined based on a trade-off between its own objectives and higher level objectives with a priority policy depending on the level of autonomous. A no-autonomous system is a system without any targeted objectives and priorities. It has just some constrains and control from other systems of higher hierarchy. The optimize operation of a non-autonomous system is defined based on the objectives of higher level system.

Based on what is mentioned, to have a comprehensive control method for a system, we should have a multi-dimensional control system that is able to perform a specific control method in each dimension. Moreover, in each control dimension it should be aware of the hierarchy and autonomy of the system in total network. This control method is called "multi-dimension multi-hierarchy"(M-D M-H) method. The mentioned concepts, urge a new line of embedded system installed in smart things and system for future compatible technology. This research, suggests computer of things, described in the following section, for system and network level that uses "multi-dimensional multi- hierarchy "control method.

In this proposed method a "dimension" is any application based on specific software of application in which SS involves is specific type of activities. A tangible example can be a mobile that has many dimension of applications like call, chat, shopping app, surfing, remote control, etc. Hence, a SS can operate in different network of operations via different application software.

"Hierarchy" is the controlling level of the system operation in each dimension of application. A suitable intelligent system design provide a control and setup facility. In IoT setups can set smartly or by user. By setup facilities a SS can be set in any level of hierarchy and autonomy in each application. On the other hand, in each dimension of application the system should be able to work by user preference method or be controlled by the specific application provider or controlling network that is in higher level of controlling hierarchy and wants to optimize its stream of service. Also in each dimension of application, the system should be able to control its lower hierarchy level connected systems. Accordingly, a future compatible design is a system that not only can have the ability of belonging to different dimension of service and application software but also can define its own place of hierarchy in each dimension of application.

However, such system needs to be equipped with a suitable programmable unit, suitable I/O ports and wireless communication system that supports different protocols. The choice of programmable unit (e.g. CoT, PIC, FPGA, CPU, etc.) depends on the system complexity, potential number of dimensions of applications, frequency of operation, speed, number of control variables, costs, etc.. The urge for wireless communication raises from the fact that the system need to be connected to different networks of application and different level of controlling hierarchy in each dimension of application. In the next section a suggested hardware with a suitable software design for future friendly system will be introduced that can facilitate construction of aimed IoT and industry 4.0.

To have a proper view on M-D M-H strategy, there is an illustration of multi-dimension multi- hierarchy product design approach in Figure.8. Each block can be a ST, SS, or SN that has ability to be connected to any suitable network of application (dimension). The hierarchy of SS in each dimension can be define by selecting proper setting options provided in the software of CoT. Each block in each dimension can control lower level blocks if is permitted and can be controlled by higher level if the permission is given.

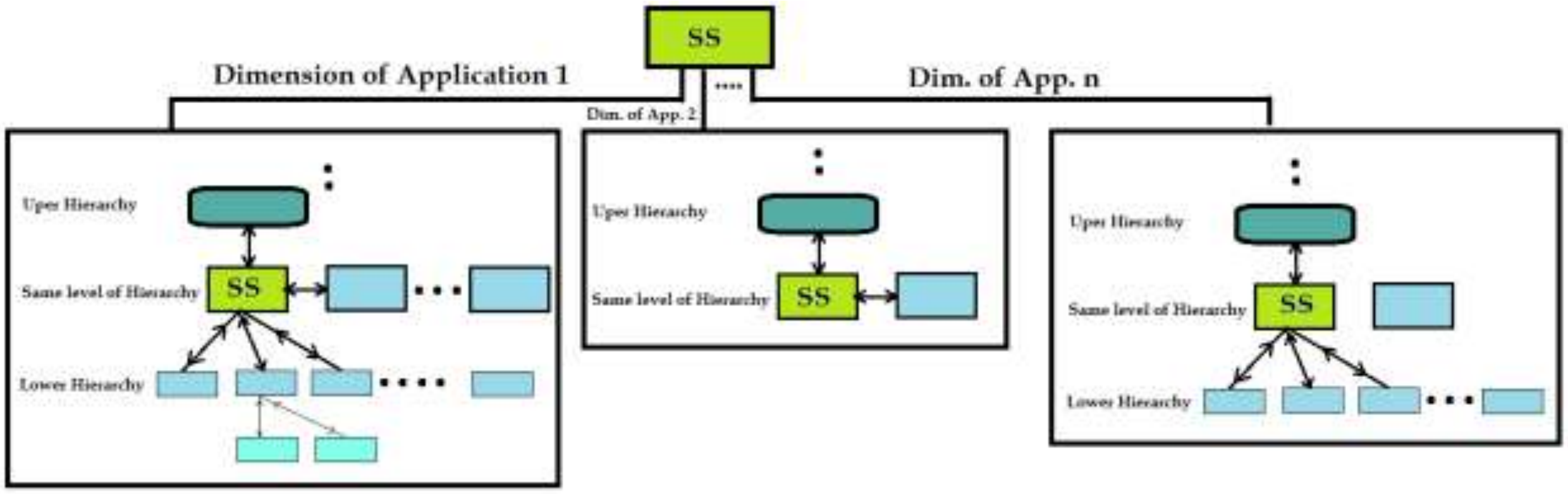

**Figure.8: Multi-dimensional multi-hierarchy (M-D M-H) control method for SSs to make it compatible with IoT. A Smart system controller shown as "SS" in multi-dimensional multi- hierarchy control scheme. In each dimension it has its own hierarchy.**

Moreover, the optimization of SS performance for each block is multi optional for each application. In each dimension or application the optimization can be; local, global, or a trade-off between local and global objectives.

## V. Proposed Computer of Things (CoTs)

There are some reasons that leads to suggestion of ready to use and classified CoTs production. First of all, manufacturers of various products (e.g. car, TV, etc.) and the owners of some organizations (e.g. hospital or an energy company) prefer to use ready-to-use and suitable packages including computer of systems and networks and software and interfaces to develop their properties for IoT rather to develop them by themselves. Many of these manufacturers or owners do not have enough information, experts or specialized R&D section for this aim. Secondly, in present time, the IoT system implementation are islanded and every leading companies are developing their systems in prototype level with their own methods using their preferred and selected communication protocols. Thirdly, there is a big market for IoT products in different classes of industries. For example car companies, TV and media companies, healthcare and hospital, agriculture, etc. Fourthly, the technology for such computer exists and many computer system producers, software developers have the capability to produce and many interface and communication protocol are developed enough to produce such suggested products. Finally, in near future, all developed systems in different classes of industries may need to interact smartly with each other. Some of this interacts may be seen in advance and some may not be predicted in time of design. Accordingly, the infrastructure of communications among SNs and SSs should be compatible with each other and can support different types of protocols. This compatibilities are in terms of physical connections, supported protocols, software support and transforming them to each other.

All mentioned reasons leads to this idea that the best way to save time, efforts, and money is to form future IoT in an organized, synchronized and classic approach. This not only needs official standardization bodies in different aspects but also needs some standards software and hardware development. Thus, CoT with different classes are suggested here. The main aim of CoT is to synchronize the development of IoT in different sector of industries using similar I/O interfaces and also communication protocols with similar or interchangeable operating systems and programming interfaces. Nowadays, smart mobile, laptops and many smart boards like UDOO, Raspberry PI, Latte panda, Pine64, wino boards, Parallella, Intel Galileo are some simple identical devices that are acting instead of proposed CoT in an islanded way. But the concept of CoT is more advanced with proper categorization in terms of type of industries and level of application.

The researchers of this paper believe that the software and hardware of CoT design should be in specific form to make it work with wide variety of products in the same class of industry and also with other classes in the standard ways in targeted IoT. As described each ST, SS, and SN need a processor with specific abilities like a computer. The processor board of each has its own power of communication, process, and storage depending on complexity of application dimensions. But as mass production of CoT will make them very cheap and affordable a high performance CoT can be used in both complex and simple systems. A CoT is very similar to a computer or smart mobile system with a big difference, in which in normal operation of this smart computer there is no human user as operator. So that we coin these types of computer by the name of "computer of thing" or CoT. CoTs should smartly set up their connections, control method and operation in any desired and possible network. A CoT should be equipped with a proper processor with some basic abilities including;

- Ability to work in different desired and possible decentralized and centralized networks of applications.
- Ability to support installation of different internet based application and support all communications protocol existed in IoT.
- Ability to operate with different communication methods existed in centralized networks. It should be able to be automatically connected to a list of SSs or STs with recognizing communication protocol in a centralized network via proper I/O port or wireless protocols. (Dynamic and automatic connectivity for location based network)
- Ability to be upgraded with different hardware and software in order to support new communication protocols in each dimension of application in both centralized and decentralized networks.
- Ability to set its hierarchy of control in each specific connection with other devices or systems both for centralized and decentralized networks.
- Ability to work in smart mode to work as plug and play system.
- Ability to work on user setting mode
- Ability to be connected to variety of sensors and actuators

To have mentioned abilities it is needed to have any ordinary product equipped with specific hardware and software design

- Proper CPU or digital controller chip for data processing
- Proper Memory
- Proper wire and wireless I/O ports such as USB, HDMI, Bluetooth, ZigBee, NFS, WiFi, cellular,
- A comprehensive RF chip and transceiver to support different wireless communication
- A proper display as user interface
- A proper operating system that can support networking, internet connection, and wide range of application software (supporting API ), etc.
- Modem and gateways and facilities for fast connections to internet depending on nature system such as DSL, WiFi, hotspot, LTE, 3G,..)
- Any other interfaces needed depending on device or system nature

It seems that a manufacturers can develop their products based on its own standards and benefits as it has been common in recent years. But the main point is that their products should be able to work with IoT and other devices and system around the world in the form of local, sub-local, sub-global, or global networks. Identical to what mobile and computer devices do with the internet structures and connectivity in order to connect and control themselves and other related devices in a local network or in a global network of internet.

Based on what is mentioned, it seems the future market urge to have new system so called “computer of things: CoT” that can provide mentioned capabilities very similar to PCs and mobiles. But it is necessary to note that the CoTs should be made for things not human as the main operator. Human just should have the ability of changing the setting and upgrading in specific time with some proper interfaces. However, as the internet of things will appear as network of networks with different layers of control, access, security and data processing, the CoTs may have various types with different capabilities from hardware and software point of view. This topic will open a new window toward new lines of research on hardware and software development for CoTs including operation system of CoT development, security methods in networks, integrated interfaces, integrated RF chips, etc.

However, in the stage of giving the Idea of IOT and before production of real CoTs that will be based on market demand priorities, classification is just speculation and hypothesis. But they would have types for system or network level, branch of industries, or based on mobile or fixed position operation.

Finally, CoTs of various types will make it very easy to make connections between different systems in both centralized and decentralized networks as they use similar communication protocols with similar hardware and software patterns in industry 4.0 and IoT era. It is highly suggested to chip designers and computer market leaders to work on mentions CoTs with different classes with variety and capabilities. This will make it easier for device and system manufacturers and also organization owners to improve their systems with ready to use and programmable packages with related interfaces available in the market. This will pave the path for future product to emerge sooner using unified standards in accordance with CoTs and other emerging supplementary equipment. The new market for CoT will emerge so the early bird will stay as leader in this market.

# VI. Conclusion

This work was concentrated on promotion of things and system towards IoT network. It tried to give manufacturer a suitable and simple outlook in order to change their productions in direction of IoT evolution. In this regard, first the need for an internationally recognized and authorized organization was discussed that can motivate governments to investigate on IoT in case, they can assure the autonomy and security of data based on locality of information. Then some standard terms including ST, SS, and SN were defined to make a simple and unanimous understanding of IoT and industry 4.0 concepts for manufacturers. Moreover, from smart system and network point of view, that means the concerns of manufacturers, future possible IoT networks were divided into two main categories including centralized and decentralized networks that are mostly based on application software. Then, it was shown that for each SS or SN to operate properly in both networks, specific hardware and software are required. Based on given speculation, the need for CoTs was predicted and a new line of design and production was proposed. CoTs would ease and accelerate establishment of IoT in unanimous and standard way by enabling systems to smartly operate in IoT network. Moreover, a brief specifications for CoTs were presented that can lead to a new line of R&D, standardization, and production in computer and chip industries. Meanwhile, based on various control hierarchy levels and also various dimension of application that a CoT would be involved, a multi-hierarchy and multi-dimensional control strategy was proposed.